\documentclass[useAMS,usenatbib,usegraphicx]{mn2e}

\title[On the Doppler boosting in J1026+2542]{On the Doppler boosting in the compact radio jet of the distant blazar J1026+2542 at $z$=5.3}

\author[S. Frey et al.]{S\'andor Frey$^{1}$\thanks{E-mail:
frey@sgo.fomi.hu}, Judit O. Fogasy$^{2}$, Zsolt Paragi$^{3}$ and Leonid I. Gurvits$^{3,4}$\\
$^{1}$F\"OMI Satellite Geodetic Observatory, P.O. Box 585, H-1592 Budapest, Hungary\\
$^{2}$Department of Astronomy, E\"otv\"os Lor\'and University, P.O. Box 32, H-1518 Budapest, Hungary\\
$^{3}$Joint Institute for VLBI in Europe, Postbus 2, 7990 AA Dwingeloo, The Netherlands\\
$^{4}$Department of Astrodynamics and Space Missions, Delft University of Technology, 2629 HS Delft, The Netherlands}

\begin{document}

\date{Accepted 2013 February 8. Received 2013 January 9; in original form 2012 November 4}

\pagerange{\pageref{firstpage}--\pageref{lastpage}} \pubyear{2013}

\maketitle

\label{firstpage}

\begin{abstract}
Based on its broad-band spectral energy distribution, and the X-ray spectrum in particular, the radio-loud active galactic nucleus (AGN) SDSS~J102623.61+254259.5 (J1026+2542) has recently been classified as a blazar. The extremely high redshift of the source, $z$=5.3, makes it one of the most distant and most luminous radio-loud AGN known to date. From published 5-GHz very long baseline interferometry (VLBI) imaging data obtained in 2006, the source has a typical blazar appearance on mas scales, with a prominent one-sided jet extending to $\sim$20~mas. We estimate the brightness temperature of J1026+2542 and find no strong evidence for Doppler boosting. The jet viewing angle is possibly at least $\sim$20$\degr$. The bulk Lorentz factor and the viewing angle of the jet could reliably be determined in the near future from multi-epoch VLBI observations.  
\end{abstract}


\section{Introduction}

Blazars are a prominent class of active galactic nuclei (AGN). Based on their optical properties, they are traditionally divided into BL Lacertae (BL Lac) objects and flat-spectrum radio quasars (FSRQ). BL Lacs have no strong emission features in their optical spectra, while FSRQs show strong, broad emission lines. Blazars are all compact radio sources with flat radio spectra (with spectral index $\alpha$$\geq$$-0.5$, defined as $S\propto\nu^{\alpha}$, where $S$ is the flux density and $\nu$ the frequency), rapid variability, high polarization, and radio structure dominated by a compact ``core''. According to the unified picture of AGN \citep[e.g.][]{Urry95}, one of their antiparallel jets points close to the observer's direction. The jets are nearly perpendicular to the accretion disk which supplies the central supermassive (up to $\sim$$10^9$~M$_\odot$) black hole with fresh material. The radiation of the relativistic outflow in the approaching jet is beamed towards us, while the receding jet becomes too faint to be detected. With the high angular resolution provided by very long baseline interferometry (VLBI), characteristic signatures of low-inclination relativistic jets, such as apparent superluminal motion and high brightness temperatures, even in excess of the inverse-Compton limit \citep[$\sim$$10^{12}$~K,][]{Kell69}, can be detected.    

Since their jets are oriented close to the line of sight, the emission of blazars is believed to be greatly enhanced by the effect of Doppler boosting. The non-thermal emission of the jet dominates the entire spectral energy distribution (SED), apart from the thermal contribution from the accretion disk, the broad-line region and the host galaxy, significant in the near-infrared, optical and ultraviolet bands \cite[see e.g.][and references therein]{Giom12}. The non-thermal emission shows two humps in the SED, one at lower frequencies (synchrotron peak), and another one at higher energies, in the X-rays or $\gamma$-rays. The high-energy emission is thought to be a result of inverse-Compton scattering of the photons by interacting with the electrons accelerated in the jet. The peak frequency of the synchrotron emission varies considerably within the blazar class, from below $10^{14}$~Hz for low-synchrorton peaked (LSP) objects to above $10^{15}$~Hz for high-synchrotron peaked (HSP) sources, with an intermediate subclass in between. 

The radio-loud AGN SDSS~J102623.61+254259.5 (J1026+2542 hereafter) has recently been claimed as the second most distant blazar by \citet{Sbar12}. Its redshift is $z$=5.266 in the Sloan Digital Sky Survey (SDSS) Data Release 9 \citep[DR9,][]{Ahn12}. J1026+2542 is a prominent radio source, with several flux density measurements available in the literature between 151~MHz and 8.4~GHz frequencies. The catalogue compiled by \citet{Voll10} gives $\alpha$=$-0.4$ for the fitted spectral index. The source is unresolved ($<5\arcsec$) at 1.4~GHz in the Karl G. Jansky Very Large Array (VLA) Faint Images of the Radio Sky at Twenty-centimeters (FIRST) survey\footnote{\tt{http://sundog.stsci.edu}} \citep{Whit97}, with $S_{\rm 1.4}$=239~mJy integrated flux density. Its total (single-dish) flux density at 4.85~GHz is $S_{\rm 4.85}$=142~mJy in the GB6 catalogue \citep{Greg96}. At this latter frequency, the source has been imaged with the US National Radio Astronomy Observatory (NRAO) Very Long Baseline Array (VLBA), as part of the VLBA Imaging and Polarimetry Survey (VIPS) sample \citep{Helm07}. 

\citet{Sbar12} studied a sample of potential high-redshift blazars selected from the SDSS, and followed up their best candidate, J1026+2542, at different wavebands to construct a SED from the radio up to the $\gamma$-rays. The syncrotron peak is located above $10^{15}$~Hz in the rest frame of the source. The blazar nature of the object is supported by its strong and hard X-ray emission and a comparison with the remarkably similar SED of the highest-redshift known blazar J0906+6930 at $z$=5.47 \citep{Roma04}. They fitted a one-zone synchrotron and inverse-Compton model, together with an accretion disk, a dusty torus and a hot thermal corona to the observed SED, and concluded that it is consistent with what is usually applied for other powerful blazars. In particular, they obtained $\Gamma$=14 for the bulk Lorentz factor in the outflow. From the thermal near infrared and optical emission, they deduced that the mass of the central black hole is between 2 and 5 billion solar masses \citep{Sbar12}.    

The number of radio AGN at very high redshifts ($z$$>$4.5) that are presently known and studied in some details, like J1026+2542, barely exceeds 10 \citep[e.g.][]{Frey10,Frey11,Frey12}. These rare objects are therefore valuable targets for observational astrophysics and cosmology. They provide important constraints on the growth of the earliest supermassive black holes and the onset of AGN activity period in the radio. Intriguingly, this small highest-redshift radio AGN sample is dominated by objects different from the highly Doppler-boosted, compact, flat-spectrum blazars. While there are several weaker (mJy-level), compact steep-spectrum radio quasars known at $z$$>$5.7, the redshift of the most distant blazar, J0906+6930 \citep{Roma04}, is only $z$=5.47. This is in contrast with the expectation that the most luminous (i.e. Doppler-boosted) sources are to be discovered first in a flux-limited sample at the highest redshifts. However, the very small sample we know at present is in fact not radio-selected and not flux-limited, but rather determined by the availability of optical spectroscopic data used to derive the redshift. It is therefore too early to conclude that blazars are ``missing'' at $z$$\ga$6.

In this paper, we re-analyse the VIPS data \citep{Helm07} taken in 2006, present an image of J1026+2542 and model its brightness distribution. This allows us to estimate the brightness temperature and the inferred Doppler-boosting factor for the inner radio jet. We present an archival 8.4-GHz VLA image to show that the source is compact at the angular scale of $\sim$500 milliarcseconds (mas). By analysing another archival VLA data set at 43 GHz, we add a new flux density point to the SED of J1026+2542. We assume a flat cosmological model with $H_{\rm 0}$=70~km~s$^{-1}$~Mpc$^{-1}$, $\Omega_{\rm m}$=0.3, and $\Omega_{\Lambda}=$0.7. In this model, $1\arcsec$ angular size corresponds to 6.12~kpc projected linear size at $z$=5.266, the luminosity distance of the source is $D_{\rm L}$=49.559~Gpc, and the age of the Universe is just above 1~Gyr \citep{Wrig06}.

\section{Radio interferometric data for J1026+2542}

\begin{table*}
  \centering 
  \caption[]{Parameters of the fitted Gaussian models for J1026+2542 at $\nu$=4.85~GHz. The statistical errors are estimated according to \citet{Foma99}, with additional 5\% flux density calibration uncertainties.}
  \label{modelfit}
\begin{tabular}{ccccc}        
\hline                 
Flux density & \multicolumn{2}{c}{Relative position } & \multicolumn{2}{c}{Component axes (FWHM)} \\
$S$ (mJy)        & north (mas) & east (mas)               & $\vartheta_1$ major (mas)     & $\vartheta_2$ minor (mas)                      \\
\hline                       
34.4$\pm$1.8 &  ...             &  ...                &  1.94$\pm$0.02  & 0.34$\pm$0.01   \\
11.3$\pm$0.7 &  1.33$\pm$0.01   &  -3.16$\pm$0.01     &  1.03$\pm$0.03  & 1.03$\pm$0.03   \\
13.2$\pm$0.9 &  1.22$\pm$0.04   &  -5.82$\pm$0.04     &  2.03$\pm$0.08  & 2.03$\pm$0.08   \\
 3.7$\pm$0.6 &  3.29$\pm$0.13   & -10.33$\pm$0.13     &  1.95$\pm$0.25  & 1.95$\pm$0.25   \\
 7.0$\pm$1.2 &  4.97$\pm$0.30   & -16.62$\pm$0.30     &  3.77$\pm$0.60  & 3.77$\pm$0.60   \\
\hline   
\end{tabular}
\end{table*}

\subsection{High-resolution VLBI image and model}
\begin{figure}
\centering
  \includegraphics[bb=70 27 520 768, height=80mm, angle=270, clip=]{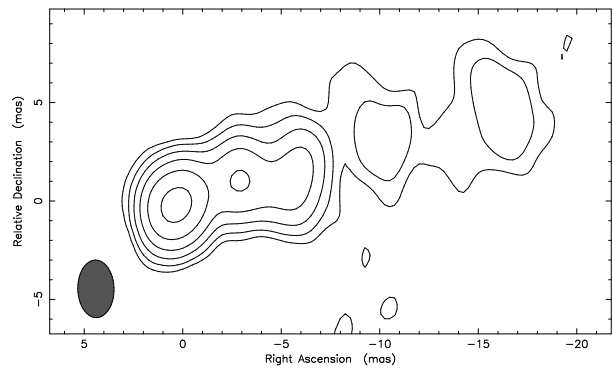}
  \caption{
The naturally-weighted 4.85-GHz VLBA {\sc CLEAN}-component image of the quasar J1026+2542 from the VIPS data taken and calibrated by \citet{Helm07}. The observing date is 2006 January 28. The lowest contours are drawn at $\pm0.6$~mJy~beam$^{-1}$ ($\sim$3$\sigma$ image noise level). Further positive contour levels increase by a factor of 2. The peak brightness is 24.2~mJy~beam$^{-1}$. The Gaussian restoring beam is 2.9~mas $\times$ 1.8~mas with major axis position angle $1\degr$. It is indicated with a filled ellipse (full width at half maximum, FWHM) in the lower-left corner.}
  \label{VLBI-image}
\end{figure}

While \citet{Sbar12} gave an account of the total flux density measurements of J1026+2542 at different radio frequencies available in the literature, they did not refer to the VLBI imaging observations by \citet{Helm07}. The image, as well as the calibrated VLBA visibility data in FITS format are publicly available from the VIPS web site\footnote{\tt{http://www.phys.unm.edu/{\textasciitilde}gbtaylor/VIPS}}. The observations of J1026+2542 were made on 2006 January 28 in 10 separate $\sim$1-min scans with the ten 25-m antennas of the VLBA. The details of the observing and the initial data calibration in the NRAO Astronomical Image Processing System \citep[{\sc AIPS},][]{Diam95} are given by \citet{Helm07}. We downloaded the calibrated visibility data and made a total intensity image of the source. We used the Caltech {\sc Difmap} package \citep{Shep94} for the conventional hybrid mapping procedure involving several iterations of {\sc CLEAN}ing and phase (then amplitude) self-calibration. Our image shown in Fig.~\ref{VLBI-image} is fully consistent whith that of \citet{Helm07}. Note that \citet{Helm07} did not detect polarized emission in this source. 

\begin{figure}
\centering
  \includegraphics[bb=70 27 520 768, height=80mm, angle=270, clip=]{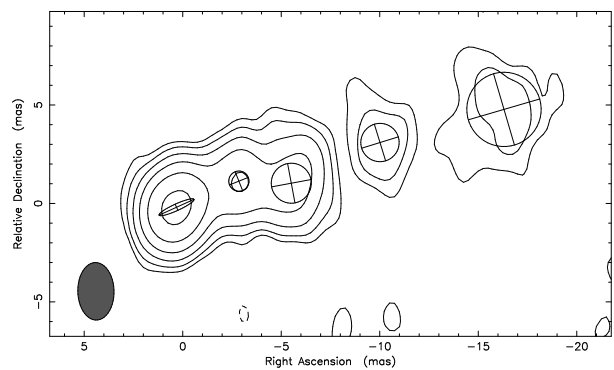}
  \caption{
The 4.85-GHz VLBA image of the quasar J1026+2542, restored with the five Gaussian model components fitted to the self-calibrated visibility data (Table~\ref{modelfit}). The lowest contours are drawn at $\pm0.6$~mJy~beam$^{-1}$. Further positive contour levels increase by a factor of 2. The peak brightness is 23.8~mJy~beam$^{-1}$. The Gaussian restoring beam is 2.9~mas $\times$ 1.8~mas with major axis position angle $1\degr$. The location and the diameter (FWHM) of the individual Gaussian model components are indicated with an ellipse for the core {\em (left)}, and circles for the extended jet. In the latter cases, the two respective axes represent directions parallel and perpendicular to the position angle of the corresponding component with respect to the core.}
  \label{VLBI-model}
\end{figure}

Our aim was to characterise the brightness distribution of J1026+2542 with a set of simple Gaussian model components. We used {\sc Difmap} for fitting a model directly to the self-calibrated visibility data. It consists of an elliptical Gaussian component for the VLBI ``core'' and four additional circular Gaussian components for the extended jet. Our model parameters are listed in Table~\ref{modelfit} and displayed in Fig.~\ref{VLBI-model}, along with the image restored using this model, which is remarkably similar to the {\sc CLEAN}-component image in Fig.~\ref{VLBI-image}. We attempted to keep the number of model parameters at the minimum, but we had to choose an elliptical over a circular Gaussian component for the ``core'' as it gave a better fit. The major axis of this elliptical points to the jet direction.

To quantitatively classify the morphology of their large sample of VIPS sources, \citet{Helm07} used an automated Gaussian model fitting procedure within {\sc AIPS}, but in the image plane. We chose visibility-plane model fitting because it is thought to be more reliable \citep{Pear95}. \citet{Helm07} also found five model components to describe J1026+2542, but listed only their mean radius (6.8~mas) and maximum separation (20.8~mas). It is obvious from Table~\ref{modelfit} that our model components fitted directly to the visibility data are smaller. 

To investigate the possibility that the central elliptical Gaussian model component represents a barely resolved blend of a weaker, more compact ``core'' and more extended blob further along the jet, we fitted and alternative model in {\sc Difmap}. The elliptical Gaussian component was replaced with a point source plus a circular Gaussian component. As the model-fitting solutions are not unique \citep{Pear95}, it is not surprising that the resulting 6-component model leads to a similarly good fit to the visibility data, and a restored image practically indistinguishable from the one in Fig.~\ref{VLBI-model}. In this representation, the first component in Table~\ref{modelfit} splits into a central point source with (15.1$\pm$0.8)~mJy flux density, and a circular Gaussian component of (17.9$\pm$1.0)~mJy flux density and (0.52$\pm$0.02)~mas angular size. Owing to the finite resolution of the VLBI array, there is a minimum resolvable angular size \citep[e.g.][]{Kova05} in this experiment, which gives an upper limit for the angular extent of the central point source, $\vartheta_1$=0.29~mas and $\vartheta_2$=0.18~mas.      

In the discussion which follows in Sect.~\ref{discussion}, we apply both the 5-component model (Table~\ref{modelfit}) and the 6-component version, with the unresolved central source, to estimate the parameters of the radio jet. In any case, according to the VLBI image and model(s), the radio structure of J1026+2542 is clearly resolved at mas scale and has a prominent one-sided core--jet structure. The jet extends up to $\sim$20~mas in the W-NW direction (position angle about $-70\degr$). The sum of the flux densities in the VLBI components (70~mJy) is the half of the total GB6 flux density measured at the same frequency, although at a different epoch. Therefore the effect of any flux density variablity, if present in the source, cannot be excluded.

\subsection{Archival VLA data}

\begin{figure}
\centering
  \includegraphics[bb=67 176 508 619, height=70mm, angle=270, clip=]{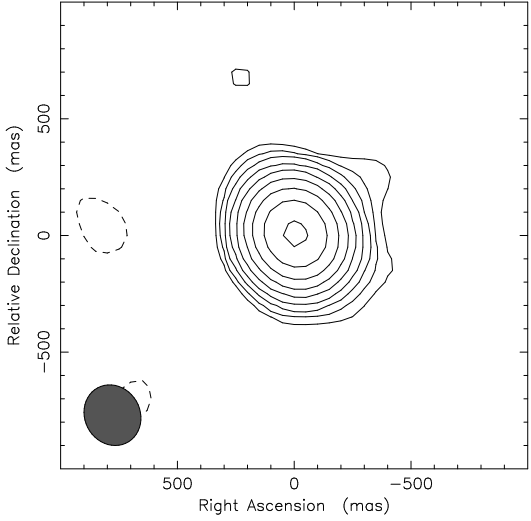}
  \caption{
The 8.4-GHz VLA A-array image of the quasar J1026+2542 (experiment AP204, 1991 June 19). The lowest contours are drawn at $\pm0.35$~mJy~beam$^{-1}$ ($\sim$3$\sigma$ image noise level). Further positive contour levels increase by a factor of 2. The peak brightness is 102~mJy~beam$^{-1}$. The Gaussian restoring beam is 266~mas $\times$ 237~mas with major axis position angle $28\degr$.}
  \label{VLA-image}
\end{figure}

At lower resolution, VLA data of the quasar J1026+2542 are available from the NRAO Archive\footnote{\tt{http://archive.nrao.edu}}. The catalogue of the CRATES 8.4-GHz survey of flat-spectrum radio sources \citep{Heal07} describes J1026+2542 as a point source with $S_{\rm 8.4}$=105.7~mJy but presents no image. We obtained archival data taken on 1991 June 19 with the VLA in its most extended A configuration (experiment code AP204). As a part of a large sample, J1026+2542 was observed for a short scan of 1.5-min duration with a total bandwidth of 100~MHz. Phases and amplitudes were calibrated in a standard way in {\sc AIPS}, using the source 3C~286 as the absolute flux density calibrator. The data were then exported to {\sc Difmap} for imaging. The 8.4-GHz image (Fig.~\ref{VLA-image}) indeed shows a source practically unresolved, with no significant radio emission beyond $0\farcs5$ above the brightness limit of $\sim$0.6~mJy~beam$^{-1}$ (5$\sigma$), although from the slight asymmetry of the contours (i.e. the shallower gradient of the brightness distribution), there is a hint on an extension towards the west. This would be consistent with the jet position angle seen in the higher-resolution VLBI image (Fig.~\ref{VLBI-image}). A circular Gaussian model component fitted to the 8.4-GHz VLA visibility data in {\sc Difmap} gives (104$\pm$5) mJy flux density, similar to the value presented by \citet{Heal07}.      

\begin{figure}
\centering
  \includegraphics[width=60mm, angle=270, clip=]{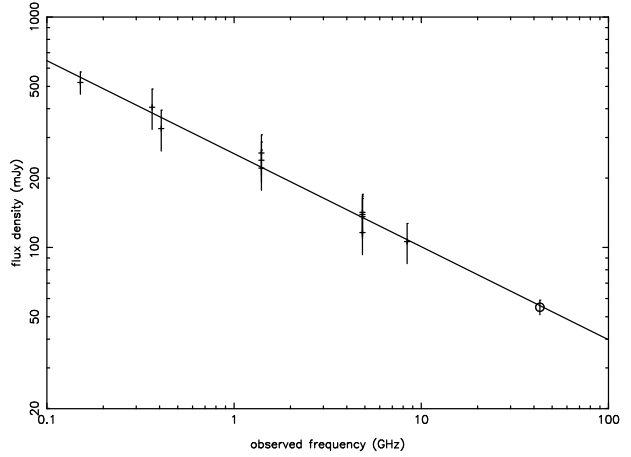}
  \caption{
The radio spectrum of J1026+2542. The total flux density data are from the catalogue compiled from the literature by \citet{Voll10}, supplemented by the 151-MHz measurement from \citet{Wald96}. Our new spectral point at the high-frequency end (43~GHz) is marked with an open circle. The line indicates the fitted power-law spectrum with $\alpha$=$-0.4$.
}
  \label{radio-spectrum}
\end{figure}

Another archival VLA A-array experiment (project code TC001, observing date 2002 June 19, bandwidth 100~MHz, on-source time 2 min, angular resolution $\sim$50~mas) allowed us to measure the flux density of J1026+2542 at 43~GHz as well. With an analysis similar to that of the experiment AP204 above, we obtained $S_{\rm 43}$=(55$\pm$4)~mJy. This indicates that the radio spectrum of the source follows the same power law with $\alpha$=$-0.4$ spectral index, from the observed frequency of 151~MHz up to at least 43~GHz (Fig.~\ref{radio-spectrum}). Due to the high redshift, these frequencies correspond to $\sim$1~GHz and 270~GHz, respectively, in the rest frame of the source. 

\section{Jet parameters from the VLBI data}
\label{discussion}

The fitted Gaussian model parameters of the VLBI ``core'' allow us to determine the brightness temperature of the source, in the units of K:
\begin{equation}
T_{\rm B} = 1.22\times 10^{12} (1+z) \frac{S}{\vartheta_1 \vartheta_2 \nu^2},
\end{equation}
where $S$ is the flux density (Jy), $\vartheta_{1}$ and $\vartheta_{2}$ are the major and minor axes of the fitted Gaussian model component (FWHM, mas).  The observing frequency, $\nu$, is expressed in GHz. 
The parameters of our resolved elliptical Gaussian ``core'' component (the first line in Table~\ref{modelfit}) lead to $T_{\rm B}$=$(1.70\pm0.13) \times 10^{10}$~K. On the other hand, if we use our point-source model from the 6-component modelfit, we obtain $T_{\rm B} > 9.4 \times 10^{10}$~K. This latter value is a lower limit to the brightness temperature due to the finite resolution of the radio interferometer. 

Let us first investigate the case of the resolved VLBI ``core'', resulting from our first model fitting (Table~\ref{modelfit}).
If we assume that the intrinsic brightness temperature ($T_{\rm B,int}$) of the source is close to the equipartition value, $T_{\rm eq} \simeq 5\times 10^{10}$~K \citep{Read94}, we can derive the Doppler factor $\delta = T_{\rm B} / T_{\rm B,int} \simeq 0.34$ \cite[see e.g.][and references therein]{Vere10}. 
Note that in at least one case, for the only very-high-redshift blazar for which the radio jet parameters have been determined \citep[J1430+4204 at $z$=4.72,][]{Vere10}, the true intrinsic brightness temparature was in good agreement with the equipartition value. 

\begin{figure}
\centering
  \includegraphics[width=65mm, angle=270]{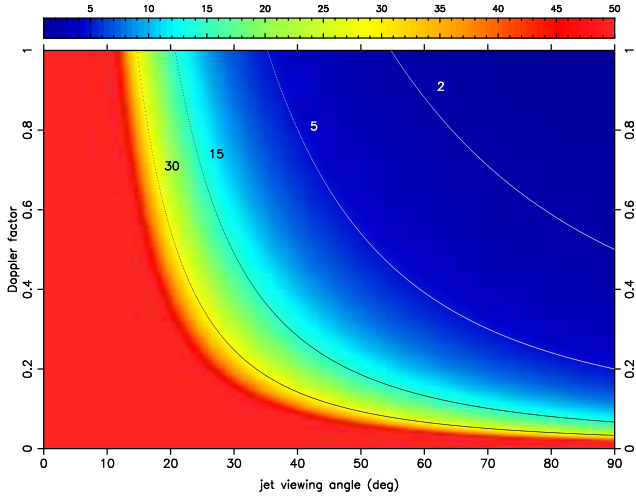}
  \caption{
The Doppler factor for the approaching jet in the deboosting regime, in the range of 0$<$$\delta$$\leq$1, as a funcion of the jet viewing angle with respect to the line of sight ($\theta$). Different values of the bulk Lorentz factor are indicated in the range 1$<$$\Gamma$$<$50; $\Gamma$=2, 5, 15 and 30 are also marked with curves.}
  \label{doppler}
\end{figure}

The small Doppler factor of $\delta$$<$1 here actually indicates Doppler {\em deboosting}, which remains the case even if we take a somewhat lower value of $T_{\rm B,int}= 2 \times 10^{10}$~K suggested by \citet{Kell04}. Doppler deboosting, i.e. the attenuation of the jet emission commonly happens in the receding jet, where $\delta$$\ll$$1$. Its effect is so severe that the AGN counterjets mostly remain invisible in VLBI images. For any given bulk Lorentz factor, there is a limiting jet viewing angle where $\delta$ falls below unity in the approaching jet as well (Fig.~\ref{doppler}).    

Using the formulae for the relativistic beaming \citep[e.g.][]{Urry95} and the presence of Doppler deboosting in J1026+2542, we can give an estimate of the jet viewing angle with respect to the line of sight. For this, we assume $\Gamma$=15 for the bulk Lorentz factor. This is an average value in a broad distribution with $\Gamma$ up to $\approx$30 found for a sample of about 30 quasars by \cite{Kell04}. We note that \citet{Sbar12} derived $\Gamma$=14 for J1026+2542 form their SED modelling. However, following the model of \cite{Ghis09}, they assumed that the jet viewing angle is $\theta$$\sim$$1/\Gamma$, therefore their proposed small viewing angle and the estimated Lorentz factor are not independent quantities.

The Doppler factor is related to the bulk Lorentz factor and the bulk speed of the jet material ($\beta$$<$1, expressed in the unit of the speed of light) as follows:
\begin{equation}  
\delta  = \frac{1}{\Gamma(1-\beta \cos \theta)}.
\end{equation}   
The Lorentz factor is
\begin{equation} 
\Gamma=(1-\beta^2)^{-\frac{1}{2}}.
\end{equation} 
Consequently,
\begin{equation} 
\cos\theta=\frac{\Gamma\delta-1}{\Gamma\delta\sqrt{1-\Gamma^{-2}}}.
\label{angle}
\end{equation} 

For $\delta$=0.34 and $\Gamma$=15, we obtain $\theta$$\approx$$36\degr$. The jet inclination angle corresponding to $\delta$=1 is $\theta$$\approx$$21\degr$. As it is apparent from Fig.~\ref{doppler}, Doppler deboosting takes place at higher jet inclination angles for lower values of Lorentz factor. On the other hand, if we consider an extreme Lorentz factor, $\Gamma=30$, close to the highest values measured e.g. by \cite{Kell04}, the viewing angles are still larger than $\theta$$\approx$$15\degr$ for Doppler deboosting to happen.

VLBI imaging observations at multiple epochs are potentially useful to measure the apparent proper motion of the components in AGN radio jets. Once the Doppler-boosting factor and the apparent transverse speeed
\begin{equation}
\beta_{\rm app}= \frac{\beta \sin \theta}{1-\beta \cos \theta},
\end{equation}
are measured, both $\Gamma$ and $\theta$ can be determined from the radio interferometric data alone. The blazar J1026+2542 is unique in the distant Universe at $z$$>$5 with its prominent mas-scale jet structure (Fig.~\ref{VLBI-image}), for which the proper motions with respect to the existing first-epoch data \citep{Helm07} could be measured now. Based on the suspected Lorentz factor and viewing angle, the expected displacement over a $\sim$7-yr interval exceeds 1~mas, reliably detectable with a second-epoch 5-GHz VLBI imaging observation. The source therefore offers the first-time possibility to directly measure AGN radio jet parameters at $z$$>$5, to compare them with known samples of lower-redshift AGN, and to supplement the data available for a classical cosmological test, the apparent proper motion--redshift relation \citep[e.g.][]{Kell04} with measurements at the extreme tail of the redshift distribution.

It is possible, however, that in the calculations above, we underestimate the brightness tempearture of the VLBI ``core'' because its emission is blended with a recently ejected, more extended jet component. Follow-up VLBI imaging observations, possibly revealing a moving inner jet component, would be essential to decide if it was indeed the case in 2006. To investigate the consequences of a higher brightness temperature, we can apply the lower limit, $T_{\rm B} > 9.4 \times 10^{10}$~K, implied by the parameters of the central unresolved source in our 6-component model fit. The comparison with the equipartition brightness temperature leads to a Doppler factor of at least 2, and a jet viewing angle smaller than about $15\degr$ (Eq.~\ref{angle}).  

An alternative way to estimate the Doppler factor is to assume that the observed X-ray emission is of inverse-Compton origin. Then the measured X-ray flux and the radio properties (optically thin spectral index, core flux density and diameter) of the source give the Doppler factor, $\delta_{\rm IC}$ \citep{Ghis93,Guij96}. Taking the {\em Swift} X-ray data from \citet{Sbar12} and the radio parameters obtained in this paper into account, and using the assumptions by \citet{Ghis93} and \citet{Guij96}, we get $\delta_{\rm IC}=0.2$ for J1026+2542. For this calculation, we considered the resolved elliptical Gaussian component for the core (Table~\ref{modelfit}). This value is close to $\delta_{\rm eq}=0.34$ we obtained from dividing the measured brightness temperature by the equipartition value. For their sample of $\sim$100 bright AGN, \citet{Guij96} also found that $\delta_{\rm IC}$$\approx$$\delta_{\rm eq}$, typically within a factor of $\sim$2. Although the $\delta_{\rm IC}$ estimate is that uncertain, the initial assumptions (in particular that the VLBI observations are taken at the turnover frequency) do not necessarily hold, and the value we obtained is rather a lower limit to the Doppler factor if there is some additional source of the X-ray flux in the object, this also suggests that the Doppler factor for J1026+2542 is likely below unity. Applying the upper angular size limit from our alternative point-source model, the calculation gives $\delta_{\rm IC}=0.8$. (The Doppler factor obtained this way would increase by a factor of $\sim$3 if the size is halved with respect to the upper limit we could derive from the VLBI data.)

\section{Conclusions}

By analysing archival VLBI imaging data, we estimated the brightness tempearture of the high-redshift flat-spectrum radio quasar J1026+2542 ($z$=5.266) and concluded that its emission is not necessarily strongly Doppler-boosted. Somewhat contrary to the common notion, blazar jets do not always point {\em very} close ($\theta$$\sim$1$\degr$) to the line of sight. In the case of the J1026+2542 jet, we have shown that $\theta$ may exceed $\sim$20--40$\degr$, if we accept a typical value of 15 for the bulk Lorentz factor, and a resolved VLBI ``core'' component. 
Even an extremely high value of the Lorentz factor (e.g $\Gamma$=30) does not change qualitatively our conclusion about the jet inclination.
If a similarly viable point-source model is used instead, the brightness temperature, and thus the Doppler factor becomes higher, and the viewing angle can be below $15\degr$. 

As the only AGN with a prominent mas- to 10-mas-scale radio jet structure at $z$$>$5 known to date, J1026+2542 holds a promise of a unique cosmic laboratory for studying jet physics. Repeated 5-GHz VLBI imaging observations in the near future, already initiated by our group, could lead to measurements of the kinematics of the jet components, providing a tool to reliably estimate the bulk Lorentz factor and the viewing angle of the radio jet. VLBI imaging with higher angular resolution, either involving longer interferometric baselines \citep[e.g. Space VLBI with the orbiting radio telescope on board the Russian {\em RadioAstron} satellite,][]{Kard13}, or shorter observing wavelengths, would be essential to better constrain the actual value of the ``core'' brightness temparature.

\section*{Acknowledgments}

We thank the anonymous referee for constructive suggestions.
The National Radio Astronomy Observatory is a facility of the National Science Foundation operated under cooperative agreement by Associated Universities, Inc. This work was supported by the Hungarian Scientific Research Fund (OTKA, K104539). 
This research has made use of the NASA/IPAC Extragalactic Database (NED) which is operated by the Jet Propulsion Laboratory, California Institute of Technology, under contract with the National Aeronautics and Space Administration, and the VizieR catalogue access tool, CDS, Strasbourg, France.
We acknowledge the work of the VIPS project team and the original investigators of the archival VLA experiments used for this study.

\label{lastpage}


\begin{thebibliography}{99}

\bibitem[\protect\citeauthoryear{Ahn et al.}{2012}]{Ahn12}
Ahn C.P. et al., 2012, ApJS, 203, 21

\bibitem[\protect\citeauthoryear{Diamond}{1995}]{Diam95}
Diamond P.J., 1995, in Zensus J.A., Diamond P.J., Napier P.J., eds, ASP Conf. Ser. 82, Very Long Baseline Interferometry and the VLBA. Astron. Soc. Pac., San Francisco, p. 227

\bibitem[\protect\citeauthoryear{Frey et al.}{2010}]{Frey10}
Frey S., Paragi Z., Gurvits L.I., Cseh D., Gab\'anyi K.\'E., 2010, A\&A, 524, A83

\bibitem[\protect\citeauthoryear{Frey et al.}{2011}]{Frey11}
Frey S., Paragi Z., Gurvits L.I., Gab\'anyi K.\'E., Cseh D., 2011, A\&A, 531, L5

\bibitem[\protect\citeauthoryear{Frey et al.}{2012}]{Frey12}
Frey S., Gurvits L.I., Paragi Z., Gab\'anyi K.\'E., Proceedings of Science, PoS(RTS2012)041

\bibitem[\protect\citeauthoryear{Fomalont}{1999}]{Foma99} 
Fomalont E.B., 1999, in Taylor G.B., Carilli C.L., Perley R.A., eds, ASP Conf. Ser. 180, Synthesis Imaging in Radio Astronomy II. Astron. Soc. Pac., San Francisco, p. 301

\bibitem[\protect\citeauthoryear{Ghisellini et al.}{1993}]{Ghis93}
Ghisellini G., Padovani P., Celotti A., Maraschi L., 1993, ApJ, 407, 65 

\bibitem[\protect\citeauthoryear{Ghisellini \& Tavecchio}{2009}]{Ghis09}
Ghisellini G., Tavecchio F., 2009, MNRAS, 397, 985

\bibitem[\protect\citeauthoryear{Giommi et al.}{2012}]{Giom12}
Giommi P., Padovani P., Polenta G., Turriziani S., D'Elia V., Piranomonte S., 2012, MNRAS, 420, 2899 

\bibitem[\protect\citeauthoryear{Gregory et al.}{1996}]{Greg96}
Gregory P.C., Scott W.K., Douglas K., Condon J.J., 1996, ApJS, 103, 427 

\bibitem[\protect\citeauthoryear{G\"uijosa \& Daly}{1996}]{Guij96}
G\"uijosa A., Daly R.A., 1996, ApJ, 461, 600 

\bibitem[\protect\citeauthoryear{Healey et al.}{2007}]{Heal07}
Healey S.E., Romani R.W., Taylor G.B., Sadler E.M., Ricci R., Murphy T., Ulvestad J.S., Winn J.N., 2007, ApJS, 171, 61 

\bibitem[\protect\citeauthoryear{Helmboldt et al.}{2007}]{Helm07}
Helmboldt J.F. et al., 2007, ApJ, 658, 203 

\bibitem[\protect\citeauthoryear{Kellermann \& Pauliny-Toth}{1969}]{Kell69}
Kellermann K.I., Pauliny-Toth I.I.K., 1969, ApJ, 155, L71

\bibitem[\protect\citeauthoryear{Kardashev et al.}{2013}]{Kard13}
Kardashev N.S. et al., 2013, Astron. Rep., in press 

\bibitem[\protect\citeauthoryear{Kellermann et al.}{2004}]{Kell04}
Kellermann K.I. et al., 2004, ApJ, 609, 539

\bibitem[\protect\citeauthoryear{Kovalev et al.}{2005}]{Kova05}
Kovalev Y.Y. et al., 2005, AJ, 130, 2473

\bibitem[\protect\citeauthoryear{Pearson}{1995}]{Pear95}
Pearson T.J., 1995, in Zensus J.A., Diamond P.J., Napier P.J., eds, ASP Conf. Ser. 82, Very Long Baseline Interferometry and the VLBA. Astron. Soc. Pac., San Francisco, p. 267

\bibitem[\protect\citeauthoryear{Readhead}{1994}]{Read94}
Readhead A.C.S., 1994, ApJ, 426, 51 

\bibitem[\protect\citeauthoryear{Romani et al.}{2004}]{Roma04}
Romani R.W., Sowards-Emmerd D., Greenhill L., Michelson P., 2004, ApJ, 610, L11 

\bibitem[\protect\citeauthoryear{Sbarrato et al.}{2012}]{Sbar12}
Sbarrato T. et al., 2012, MNRAS, 426, L91 

\bibitem[\protect\citeauthoryear{Shepherd et al.}{1994}]{Shep94}
Shepherd M.C., Pearson T.J., Taylor G.B., 1994, BAAS, 26, 987

\bibitem[\protect\citeauthoryear{Urry \& Padovani}{1995}]{Urry95}
Urry C.M., Padovani P., 1995, PASP, 107, 803

\bibitem[\protect\citeauthoryear{Veres et al.}{2010}]{Vere10}
Veres P., Frey S., Paragi Z., Gurvits L.I., 2010, A\&A, 521, A6 

\bibitem[\protect\citeauthoryear{Vollmer et al.}{2010}]{Voll10}
Vollmer B. et al., 2010, A\&A, 511, A53 

\bibitem[\protect\citeauthoryear{Waldram et al.}{1996}]{Wald96}
Waldram E.M., Yates J.A., Riley J.M., Warner P.J., 1996, MNRAS, 282, 779

\bibitem[\protect\citeauthoryear{White et al.}{1997}]{Whit97}
White R.L., Becker R.H., Helfand D.J., Gregg M.D., 1997, ApJ, 475, 479 

\bibitem[\protect\citeauthoryear{Wright}{2006}]{Wrig06}
Wright E.L., 2006, PASP, 118, 1711 

\end{thebibliography}
\end{document}